\documentclass[11pt]{llncs}
\usepackage{amsmath,amsfonts,amssymb,mathtools,url, paralist}
\usepackage{tikz}
\usepackage[english]{babel}

\def\XLR#1{\xleftrightarrow[\rule{2cm}{0pt}]{#1}}
\def\XR#1{\xrightarrow[\rule{2cm}{0pt}]{#1}}
\def\XL#1{\xleftarrow[\rule{2cm}{0pt}]{#1}}

\def\taxonomy{\mathcal{T}} 

\def\Zahl{\mathbb{Z}} 
\def\G{\mathbb{G}_1}
\def\GT{\mathbb{G}_T}

\def\vendor{\mathcal{V}} 
\def\part{\mathcal{P}}
\def\cust{\mathcal{C}}

\def\iseq{\stackrel{?}{=}} 

\title{Privacy-preserving Loyalty Programs} 

\author{Alberto Blanco-Justicia \and Josep Domingo-Ferrer}

\institute{Universitat Rovira i Virgili\\
Dept. of Computer Engineering and Mathematics\\
UNESCO Chair in Data Privacy\\
Av. Pa\"{\i}sos Catalans 26\\
E-43007 Tarragona, Catalonia\\
\email{\{alberto.blanco,josep.domingo\}@urv.cat}}

\begin{document}
\maketitle

\begin{abstract}
Loyalty programs are 
promoted by vendors to incentivize loyalty in buyers. 
Although such programs have become widespread,
they have been criticized by business experts
and consumer associations: loyalty results in profiling
and hence in loss of privacy of consumers.
We propose a protocol for privacy-preserving 
loyalty programs that allows
vendors and consumers to enjoy the benefits
of loyalty (returning customers
and discounts, respectively), while allowing consumers
to stay anonymous and empowering them
to decide how much of their profile they
reveal to the vendor. The vendor must offer
additional reward if he wants to learn more 
details on the consumer's profile.  
Our protocol is based on
partially blind signatures and
generalization techniques, 
and provides anonymity
to consumers and their purchases, while still
allowing negotiated consumer profiling.\\  

\end{abstract}

\section{Introduction}

Loyalty programs are marketing efforts 
implemented by vendors, especially retailers,
that are aimed at establishing a lasting
relationship with consumers.
In a loyalty program, the vendor pursues
two main goals: i) to encourage the consumer
to make more purchases in the future (returning
customer); ii) to allow the vendor to profile
the consumer in view of conducting market
research and segmentation (profiled customer). 
In order to lure
consumers into a loyalty program, the vendor
offers them rewards, typically loyalty points that
consumers can later exchange for discounts, gifts or 
other benefits offered by the vendor.
Normally, enrollment to loyalty programs involves some kind of
registration procedure, in which customers fill out a form with
their personal information and are granted a loyalty card, be it
a physical card (magnetic stripe or smartcard) or  
a smartphone application.

Market analysis and client segmentation are carried out by
building profiles of individual customers based on their personal
information, which customers supply to the vendor during
enrollment to the loyalty program, and their purchase records, 
collected every
time customers present their loyalty cards.
The profiles thus assembled are used in marketing actions,
such as market studies and targeted advertising.

Although loyalty programs have become widespread, 
they are experiencing a loss of active 
participants and they have been criticized by business experts
and consumer associations. Criticism is mainly due
to privacy issues, because it is not always clear whether the 
benefits vendors offer in their loyalty programs
are worth the loss of consumer privacy
caused by profiling~\cite{Rep1,Rep2,Rep3,Rep4}.

Loyalty programs can offer clear advantages to both
vendors and consumers, like returning customers and special discounts,
respectively.
However, privacy concerns regarding buyer profiling
affect more and more the acceptance of such programs, 
as the public awareness on the dangers of personal information
disclosure is increasing. 

\subsection{Contribution and plan of this paper}

In this work we propose a protocol for privacy-preserving
loyalty programs that allows vendors and consumers to enjoy
the benefits of loyalty, while preserving the
anonymity of consumers and empowering them to decide how accurately
they reveal their profile to the vendor.
In order to encourage customers not just to return
but also to disclose more of their
profile, the vendor must offer additional rewards to consumers.
Thus, vendors {\em pay}
consumers for their private information.
On the other hand, consumers become aware of how much
their personal data are worth to vendors, and they
can decide to what extent they are ready to reveal such data
in exchange for what benefits.

To empower consumers as described above, 
we provide them with a mechanism
that allows them to profile themselves, generalize their 
profiles and submit these generalized profiles 
to the vendor in an anonymous way.
There are some technical challenges to be overcome:
\begin{itemize}
\item The proposed mechanism should prevent vendors from linking the
generalized profiles to the identity of buyers, to particular
transactions or to particular loyalty points submitted
for redemption.
\item To prevent straightforward profiling by the vendor,
payment should be anonymous.
In online stores, to completely achieve anonymity, 
the buyers should use some kind of anonymous payment system, 
such as Bitcoin~\cite{Bitcoin}, Zerocoin~\cite{Zerocoin}, 
some other form of electronic cash~\cite{Chaum90},
or simply scratch cards with prepaid credit anonymously
bought, say, at a newsstand. In physical
stores, it would be enough to pay with cash.
\item Consumers should not be able to leverage their anonymity
to reveal forged profiles to the vendor, 
which would earn them rewards without
actually revealing anything on their real purchase pattern. 
\end{itemize} 

Our proposed mechanism, thus, needs to take care of the 
two main aspects of loyalty programs. 
First, it has to provide a way to obtain and submit loyalty points
in an anonymous and unlinkable way; that is, a customer should be able
to submit a particular loyalty point to a vendor, but the vendor 
should not be able to link that particular loyalty point to the 
transaction in which it was issued.
Second, our mechanism must allow customers to build their own 
generalized profiles from their purchase history, 
but it must prevent customers to forge false profiles and
vendors to link the generalized profiles to particular customers.
We will show later that these two aspects can be tackled
in a similar way.

The paper is organized as follows.
Section~\ref{sec.req} describes a traditional loyalty
program and presents the requirements and security 
properties our new protocol should satisfy.
In Section~\ref{sec.annt} we introduce a cryptographic
protocol based on partially blind signatures 
that is the basis of our proposed solution.
Section~\ref{sec.gene} discusses a generalization
strategy that our protocol will follow.
In Section~\ref{sec.protocol} we present our privacy-preserving
loyalty program protocol.
In Section~\ref{sec.experiments} we analyze the
performance of the system in terms of computation
and communication complexity.
Finally, Section~\ref{sec.conclusions} summarizes 
our conclusions and plans for future work.

\section{Loyalty programs}\label{sec.req}

Our method aims to offer all the functionalities
of loyalty programs; that is, to allow vendors to
reward returning customers with
loyalty points and profile 
returning customers based on their purchase histories.
The novelty is that our scheme 
empowers customers with the ability to decide 
how accurately they disclose 
their purchase histories 
to vendors. 

A simple and perhaps the most widespread approach
to implement a loyalty program is to have a centralized
server, owned and operated by some vendor $\vendor$,
that stores the information on the program participants.
This information includes all the personal data the participants
gave to the vendor when they enrolled to the program,
their balance of loyalty points, and their history of purchases.
Each customer is given a loyalty card which contains
the identifier of her record in the server's database.
Each time a customer buys at a store and
presents her loyalty card, her record in the server is
updated, by adding to it the items she bought and modifying
her balance of loyalty points if needed.
In this way, all transactions by each customer
can be linked to each other using the customer's identifier. 
Even if the customer provided false information when 
she enrolled to the loyalty program, all of her transactions
would be linked anyway. Hence, discovering the customer's identity
in one individual transaction ({\em e.g.} through the 
credit or debit card used for payment)
would allow linking her entire profile
to her real identity.

If control over profiling and purchase histories is to be 
left to customers, a centralized approach does not seem 
a good solution. Moreover, we should also ensure
that individual transactions cannot be linked to each other
unless desired by the customer.
To do so, we will let each customer manage locally
and anonymously her own balance of loyalty points
and history of purchases.

\subsection{A privacy-preserving alternative}

Our proposed mechanism follows the 
decentralized approach. To allow local management
of loyalty points and purchase receipts by the customer,
we treat points and receipts as anonymous electronic cash that
is issued by vendors and which can only be redeemed
at the vendor who issued it.
Moreover, the concrete implementation of the
loyalty program should discourage customers
from transferring loyalty points and purchase
receipts among them.
Purchase histories will be built by 
the vendor from the individual purchase
receipts of all products purchased by each customer
{\em that the customer allows the vendor to link together};
furthermore, the customer can decide
how generalized/coarsened are the product descriptions
in the purchase receipts she allows the vendor to link
to one another.

Our proposed loyalty program protocol suite consists
of the following procedures:

\begin{itemize}
\item {\sc Setup}. Algorithm run by some designated entity, 
which, on input a security parameter, outputs the parameters of the
system. These parameters can be common to several
vendors.
\item {\sc VendorSetup}. Protocol run by a vendor $\vendor$ 
in which the specific loyalty program is set up. Also,
$\vendor$ obtains the public parameters of the
system and a key pair.
\item {\sc Enroll}. Protocol run by some customer $\cust$
whereby $\cust$ is given access to the loyalty program
and the means to participate in it, typically a loyalty
card or a smartphone application.
\item {\sc Use}.  Interactive protocol run
between some $\vendor$ and $\cust$, in which $\cust$ inputs
the name of a product she wants to buy 
and obtains a 
purchase receipt which proves
that $\cust$ has purchased 
the product from $\vendor$.
\item {\sc Submit}. Interactive protocol run
between some $\vendor$ and $\cust$, in which 
$\cust$ submits a list of possibly generalized purchase receipts
to $\vendor$, in order to get loyalty points.
\item {\sc Issue}. Interactive protocol run 
between some $\vendor$ and $\cust$, in which $\cust$ 
obtains a certain amount of loyalty points.
\item {\sc Redeem}. Interactive protocol run
between some $\vendor$ and $\cust$, in which 
$\cust$ submits a certain amount of loyalty points to $\vendor$
to obtain some benefits.
\end{itemize}

\subsection{Desirable properties}
\label{desirable}

We can state the following requirements 
for loyalty points and purchase receipts: 
\begin{itemize}
\item {\bf Correctness.} Loyalty points issued by a $\vendor$
to a $\cust$ following the {\sc Issue} protocol are accepted
by the same $\vendor$ running the {\sc Redeem} protocol. 
Similarly, purchase receipts given by $\vendor$ to some
$\cust$ during the {\sc Use} protocol will be accepted by
the same $\vendor$ in the {\sc Submit} protocol,
even if they have been generalized.
\item {\bf Unforgeability.} It should be impossible to any 
malicious customer or any coalition of malicious 
customers and vendors to forge new loyalty points 
or receipts issued by a vendor $\vendor$, regardless of how many 
original loyalty points or receipts they own from $\vendor$.
\item {\bf Anonymity.} 
Loyalty points and purchase receipts should 
be granted in an anonymous way.
A vendor should be unable to learn anything about a customer 
redeeming points or submitting 
receipts, other than the customer legitimately owns them.
This should hold even if the vendor colludes with other vendors
or customers.
\item {\bf Controlled linkability.}
A customer $\cust$ should be able to decide
whether a submitted purchase receipt can
be linked to other purchase receipts
submitted by $\cust$ to the same vendor $\vendor$.
\end{itemize}


\section{Anonymous tokens with controlled linkability}\label{sec.annt}
As stated in the previous section, loyalty points and 
purchase receipts have requirements
in line with those of anonymous
electronic cash and anonymous electronic
credentials.
These well-known technologies 
use blind signatures and/or 
zero-knowledge proofs of 
knowledge~\cite{Camenisch01,Camenisch04,Chaum90,Neff01}.
We will
treat points and receipts using a construction that 
we call anonymous
tokens. 
These tokens will be realized by using 
partially blind signatures
with some additional features.

\subsection{Partially blind signatures}
Blind signature protocols are interactive protocols
between a requester and a signer, in which the signer produces 
a digital signature of a message submitted by the requester, 
but does not learn anything about the message content.
This primitive was introduced by Chaum in~\cite{Chaum83} 
and has since been used in a vast array of
privacy related protocols, such as e-cash, electronic voting
and anonymous credential systems.
An inherent drawback of blind signature protocols
is that the signer cannot enforce a certain format on
the message. Traditionally, this problem has been 
solved using {\em cut-and-choose} techniques, 
in which the requester of a signature generates and
blinds a number $n$ of messages, the signer asks
the requester to unblind all messages but a randomly
chosen one, checks
whether all unblinded messages conform to the required format and,
if yes, signs the only message that remains blinded.
Using {\em cut-and-choose} techniques solves the problem
(the probability that the requester succeeds 
in getting a non-conformant message signed
is upper-bounded by $1/n$),
but it does so at the cost of high computation and communication
overheads.

Partially blind signatures were introduced by Abe 
in~\cite{Abe96} as an alternative to {\em cut-and-choose}
protocols. In a partially blind signature protocol, the
requester and the signer agree on a public information
that is to be included in the signed message, the signer
can be sure that such information 
is really included, and the requester can be sure
that the signed message remains blinded to the signer. 

We use a partially blind signature scheme from 
bilinear pairings presented in~\cite{PBBLS}.
This scheme satisfies the requirements of completeness,
partial blindness and unforgeability against
one-more forgery under chosen message attacks, and thus
it is considered secure. Security proofs can be found in the
original paper. Additionally, this scheme produces short signatures, 
it is computationally efficient and allows aggregate 
verification of signed messages bearing 
the same agreed public information.

\subsection{Controlled linkability of tokens}
The use of partially blind signatures will ensure
that a submitted token cannot be linked
to an issued token, nor to the customer to whom
it was issued.
However, if vendors are to be allowed to build customer profiles
from anonymous purchase receipts, there must be  
a mechanism whereby, if allowed by the customer, 
the vendor can verify that several submitted purchase
receipt tokens really correspond to the same 
(anonymous) customer, even if receipts have been generalized
by the customer prior
to submission. 
Note that if all (ungeneralized) purchase 
receipts from the same customer
could be linked, 
customer anonymity would be problematic 
in spite of partially blind signatures: 
a very long and detailed profile is likely 
to be unique and goes a long way towards leaking
the customer's identity.

Thus, we propose a mechanism that allows customers 
to decide which purchase receipt tokens can be linked together,
by employing an additional identifier as part of the
secret message in the partially blind signature.
This identifier is chosen by the customer for
each receipt token at the moment of token issuance. 
If a customer picks a
fresh random number for each issued purchase receipt, then
none of this customer's receipts will be linkable to each other;
however, if the customer uses the same identifier
for a group of purchase receipt tokens at the time
of token issuance, then all of the
tokens in this group can be verifiably linked together
by the vendor after they are submitted.

%
%

\subsection{Description}
\label{desc}

Anonymous 
tokens
are operated in four phases:
\begin{itemize}
\item In the \emph{setup} phase,
a certification agency generates the public parameters
of the partially blind signature scheme. 
\item In the \emph{key generation} phase, users ({\em i.e.}
vendors and customers) get their
key pairs from the certification agency.
\item In the \emph{issuance}
phase, a token corresponding
to some loyalty points or to a purchase receipt 
is generated by a customer, it is signed in a partially blind
way by a vendor and it is returned to the customer. 
\item Finally, in the \emph{verification} phase, a customer submits
previously generated tokens to a vendor, who in turn
verifies that each token was correctly
signed. 
If tokens correspond to purchase receipts, the vendor may verify
whether the submitted tokens are linked with each other and/or
with previously submitted tokens.
\end{itemize} 

\subsubsection{Setup.} 

This algorithm is executed once by a certification authority
to set up the system parameters. It takes as input a security
parameter $\lambda$.
The algorithm chooses bilinear groups $(\G, \GT)$
of order $q > 2^\lambda$, 
an efficiently computable bilinear map 
$e: \G \times \G \rightarrow \GT$, a generator $g \in \G$
and collision-resistant hash functions 
$H: \{0,1\}^* \rightarrow \Zahl_q^*$ and 
$H_0: \{0,1\}^* \rightarrow \G$.
The public parameters are
$\mathsf{pms} = \{\G, \GT, e, q, \lambda, g, H, H_0\}.$

\subsubsection{Key generation.}

A vendor gets a secret key
$sk_{\vendor} = x \in_R \Zahl_q^*$ and a public key $pk_{\vendor} = g^x$,
and publishes his public key.

\subsubsection{Token issuance.}

\begin{figure}[!h]
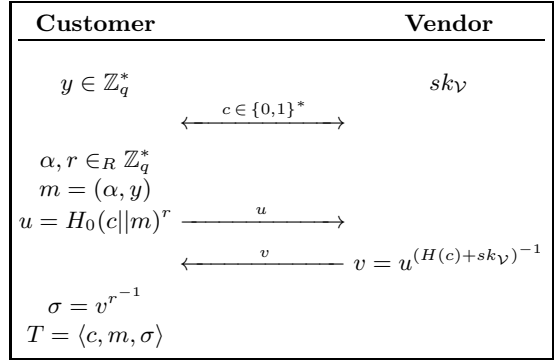

\centering
	$ \boxed{
	\begin{array}{@{} c c c @{}}
		{\bf Customer}  &  & {\bf Vendor} \\
		\hline \\
		y \in \mathbb{Z}_q^* & & sk_{\vendor} \\
		& \XLR{c \, \in \, \{0,1\}^*} \\
		\alpha, r \in_R \mathbb{Z}_q^* \\
		m = (\alpha, y) \\
		u = H_0(c||m)^r & \XR{u} \\
		& \XL{v}  & v = u^{(H(c)+sk_{\vendor})^{-1}} \\
		\sigma = v^{r^{-1}} \\
		T = \langle c, m, \sigma \rangle \\
	\end{array}
	} $
\caption{Token issuance protocol}
\label{fig.issue}
\end{figure}

A customer wants to obtain from a vendor a token with an agreed public
information $c$ (this information may specify
a number of loyalty points or a purchase receipt for a certain product). 
This is an interactive
protocol which produces a partially blind signature on
public information $c$,  
and a secret message containing a
unique identifier $\alpha$ of the token
and a (possibly) unique identifier $y$.
The protocol is depicted in Figure~\ref{fig.issue} and described
next:
\begin{enumerate}
\item The customer chooses a value for $y$, either from a list of previously
used values or by generating a new one uniformly at random from $\Zahl_q^*$.
\item The customer and the vendor agree on a public string $c \in \{0,1\}^*$.
\item The customer chooses random $\alpha, r \in_R \Zahl_q^*$ and builds
the message 
$m = (\alpha, y)$.
Then, the customer blinds the message by computing $u=H_0(c||m)^r$ and sends 
$u$ to the vendor.
\item The vendor signs the blinded message by computing 
$v =  u^{(H(c)+sk_{\vendor})^{-1}}$ and sends it back to the customer.
\item The customer unblinds the signature by computing 
$\sigma = v^{r^{-1}}$. 
The resulting tuple $T = \langle c, m, \sigma \rangle$ is the token.
\end{enumerate}

An execution of this protocol, between a vendor $\vendor$ and a customer
$\cust$, is denoted by $T = \langle c, m, \sigma \rangle = 
\mathsf{Issuance}(\vendor, \cust, c, y)$.

\subsubsection{Token verification.}
\begin{figure}[!h]
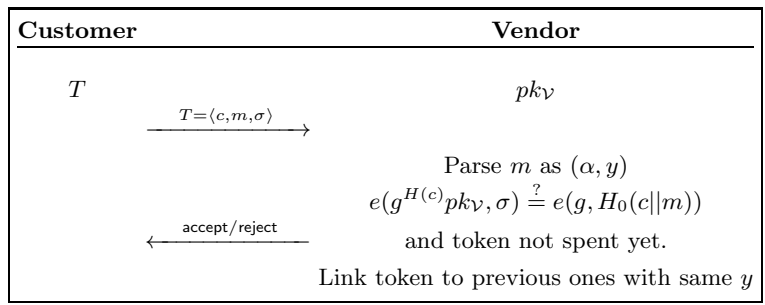

\centering
	$ \boxed{
	\begin{array}{@{} c c c @{}}
		{\bf Customer}  &  & {\bf Vendor} \\
		\hline \\
		T & & pk_{\vendor} \\
		& \XR{T = \langle c, m, \sigma \rangle} \\
		& & \text{Parse } m \text{ as } (\alpha, y) \\
		&  & e(g^{H(c)} pk_{\vendor}, \sigma) \iseq  e(g, H_0(c||m)) \\
                     & \XL{\sf accept/reject} & \text{and token not spent yet.} \\
		&	& \text{Link token to previous ones with same $y$}\\	
	 \end{array}
	} $
\caption{Verification protocol}
\label{fig.submit}
\end{figure}

The submission and verification of a token is an interactive
protocol between a customer and a vendor. The customer submits
the token $T= \langle c, m, \sigma \rangle$ and the vendor
returns {\tt accept} or {\tt reject} as a result of the
verification.
Informally, the vendor checks that the signature on the
token is valid and has been produced by himself;
then, if the value $y$ contained in the message matches the one of
a previously submitted token, the tokens are grouped.
The protocol is outlined in Figure~\ref{fig.submit} and 
described next:
\begin{enumerate}
\item The customer sends $T= \langle c, m, \sigma \rangle$ to
the vendor.
\item The vendor parses the message $m$ as 
$(\alpha, y)$.
\item The vendor verifies the signature by checking the equality
$$ e(g^{H(c)}\cdot pk_{\vendor}, \sigma) \iseq e(g, H_0(c||m)).$$
If the above equality holds, check whether the token has already been
spent (verify whether a token with the same $\alpha$ has
previously been submitted).
If the verification was successful and the token has not
been spent yet, mark it as spent and send an 
{\tt accept} message to the customer. 
Otherwise, send a {\tt reject} message to the customer.
\item Finally, the vendor checks whether the identifier
value $y$ is the same as the one in a previously spent token. If yes,
link the new token with that previous one.
\end{enumerate}

An execution of this protocol involving a customer $\cust$, 
a vendor $\vendor$ and
a token $T$ is denoted as $\mathsf{accept/reject}  = \mathsf{Verification}(\vendor, \cust, T)$.

\subsubsection{Aggregate verification.}

This protocol allows the customer to aggregate signatures
of messages bearing the same public information by just
multiplying the resulting signatures. If there is a list of
tokens $\{T_1, \dots , T_n\}$, where $T_i = \langle c_i, m_i, \sigma_i \rangle$,
and $c_i=c$ for $1 \le i \le n$, a customer can
aggregate the partially blind signatures by computing $\sigma_{agg} = \prod_{i=1}^n{\sigma_i}$ and
submitting $T_{agg} = \langle c, \{m_1, \cdots, m_n\}, \sigma_{agg} \rangle$.
The vendor can then verify the validity of the aggregated token by
checking the equality
$$ e(g^{H(c)}\cdot pk, \sigma_{agg}) \iseq e(g, \prod{H_0(c||m_i)}).$$


\subsection{Security analysis}

The desirable security features that were described in 
Section~\ref{desirable} are 
satisfied by the above protocol suite, as argued
below.
\begin{itemize}
\item {\bf Correctness.}
If the partially blind signature scheme is correctly computed, the 
verification equation will pass, because
\begin{eqnarray*}
e(g^{H(c)}\cdot pk, \sigma) &=& e(g^{H(c) + x}, \sigma) \\
&=& e(g^{H(c) + x}, v^{r^{-1}}) \\
&=& e(g^{H(c) + x}, u^{(H(c)+x)^{-1}\cdot r^{-1}}) \\
&=& e(g^{H(c) + x}, H_0(c||m)^{r\cdot r^{-1} \cdot (H(c)+x)^{-1}}) \\
&=& e(g, H_0(c||m)^{r\cdot r^{-1} \cdot (H(c) + x) \cdot (H(c)+x)^{-1}}) \\
&=& e(g, H_0(c||m)).
\end{eqnarray*}


\item{\bf Unforgeability.}
Unforgeability is provided by the partially blind signature
scheme. The security proofs can be found in the original 
work in~\cite{PBBLS}.
\item{\bf Anonymity.}
No information on the user is obtained by a server during the protocol.
Submitted tokens cannot be linked to issued tokens or to the identity
of a requester or prover because of the {\em partial blindness} property
of the signature scheme. 
\item{\bf Controlled linkability.}
When a token is issued, the identifying value $y$ is only
known to the customer who generated the token, 
due to the {\em partial blindness} of the signature.
Hence, if two verified tokens contain the same identifying value $y$,
there are two possibilities: i) both tokens were generated 
by the same customer, who re-used $y$ to allow the vendor to link them; 
ii) the customer who generated one token 
leaked $y$ to the customer who generated the other token.
If the latter leakage is prevented by technical means or discouraged
with appropriate incentives (see discussion in Section~\ref{reward} below), 
then two tokens containing the same
$y$ can be linked by the vendor as corresponding to the same customer. 
\end{itemize}


\section{Generalization of purchase histories}\label{sec.gene}

To implement our protocol, a vendor must use a publicly
available taxonomy for the products he offers.
This taxonomy $\taxonomy$ is modeled as a tree, being
its root node a generic identifier such as \emph{Product}, and each
leaf a specific product in the set of  
products $ P = \{ p_1, \dots  , p_n \}$ on sale.
The inner nodes of the tree are the subsequent categories to which the 
products belong: the closer to the leaf nodes, the more specific
categories are.
A generalization function $g: \taxonomy \rightarrow \taxonomy$ 
returns the parent of a node. 
Applying the generalization function $m$ times will
be denoted as $g^m$.
As an example, for the product $p_i = Inception$, its generalizations
might be $g(p_i) = ActionMovie$, $g^2(p_i)=Movie$, $g^3(p_i)=DigitalMedia$
and $g^4(p_i)=Product$.
For simplicity and ease of implementation,
it is desirable that all leaves be at the same depth,
that is, that the path from the root to any leaf be of the same 
length. 

Customers in our loyalty program protocol will receive a
list of anonymous tokens, each issued as described in the 
previous section, for every product they purchase.
This list contains a receipt for the specific product and
receipts for all of its generalizations in the path up to the root 
of the taxonomy (generalization path).
When a customer decides to submit her purchase history,
she chooses the level of generalization she wants 
for each purchase. Then the customer sends 
for each purchase the tokens in the 
purchase generalization path from the chosen generalization
level up to the root of the taxonomy. 
Following the movie example above,
a customer who wants to submit her purchase
generalized to level 2 will submit the tokens $Movie$, 
$DigitalMedia$ and $Product$.
Forcing customers to send all tokens from
the selected generalization level to the root
prevents them from using tokens in the generalization
path of a purchase to falsely claim other purchases.


\section{Privacy-preserving loyalty program construction}\label{sec.protocol}


Our proposed solution for privacy-preserving loyalty
programs builds on the anonymous 
tokens we described
in Section~\ref{sec.annt} and the generalization of purchase
histories described in Section~\ref{sec.gene}.
As introduced in Section~\ref{sec.req}, our construction
consists of the following protocols:
\textsc{Setup, VendorSetup,
Enroll, Use, Submit, Issue} and {\sc Redeem}.

\subsection{Setup}
The setup phase is run by a certification authority to generate
the public parameters \textsf{pms} of the anonymous
token construction described in Section~\ref{sec.annt}.
The system parameters are made public to every $\vendor$ offering
loyalty programs and to every  
$\cust$ intending to participate in them.

\subsection{VendorSetup}

Each vendor $\vendor$ publishes a product taxonomy 
$\taxonomy_{\vendor}$ as des\-cribed in 
Section~\ref{sec.gene}.
Then, $\vendor$ obtains a key pair built 
as described in the key generation procedure in 
Section~\ref{sec.annt}. 
Finally, $\vendor$ publishes his public key.

\subsection{Enroll}
Customers obtain the public parameters of the system and some
means to communicate with the system, namely a smartcard or
a smartphone application.
Furthermore, customers enrolling to a loyalty program from a particular 
vendor obtain the vendor's public key and his taxonomy of products. 
This step is not mandatory, but it allows customers to check 
that tokens issued by vendors are valid and purchase
receipt generalizations
are correct. 

\subsection{Use}

A customer $\cust$ in a loyalty program offered by
a vendor $\vendor$ 
purchases a product, either at
a physical or online store of $\vendor$. Note that,
in the case of an online store, $\cust$ should
use additional anonymization measures, such
as anonymous Internet surfing, offered for example
by Tor networks~\cite{ToR}, anonymous shipping
methods~\cite{eDropship}, and anonymous payment
methods ({\em e.g.} \cite{Bitcoin,Chaum90,Zerocoin}
or simply prepaid scratch cards). The protocol is as follows:

\begin{enumerate}
\item $\cust$ sends to $\vendor$ the name $p_i$ of the product
$\cust$ wants to buy. 
\item $\cust$ chooses a value $y$ to be used in 
the token issuance protocol, depending
on her privacy preferences: if she wants the new purchase receipt to be 
linkable to
previously obtained purchase receipts (linkability 
is incentivized as described in Section~\ref{reward} below), she will re-use the same $y$ that
was used in those previous receipts; 
if she does not want this new purchase receipts to be linkable
to previous receipts, she will pick a new random $y \in \mathbb{Z}^*_q$.
\item In order to produce purchase receipt tokens
for product $p_i$ and all its generalizations, $\vendor$ and $\cust$ run 
the interactive protocol $\mathsf{Issuance}(\vendor, \part, p_i, y)$,
$\mathsf{Issuance}(\vendor,  \part, g(p_i), y)$,
$\mathsf{Issuance}(\vendor, \part, g^2(p_i), y)$, etc. up to the root
of the taxonomy. In this way, 
$\cust$ obtains as many purchase receipt tokens as the depth
of $p_i$ in $\vendor$'s taxonomy.
\end{enumerate}

\subsection{Submit} 

At any moment, a customer can submit a list of 
purchase receipts (or a generalized version of them) to the
vendor and obtain loyalty points. To this end,
for each purchased product in her claimed purchase history, 
the customer sends the receipt token corresponding
the level of generalization 
she wishes. Additionally, for each product, she also submits
all tokens from the selected 
generalization level up to the root of the taxonomy
(to make sure tokens in the generalization path cannot
be later used as independent purchase receipts). The submission
of each token $T_i$ is performed 
according to the $\mathsf{Verification}(\vendor, \part, T_i)$
protocol described in Section~\ref{desc}.

\subsection{Issue} 

To issue loyalty points, the vendor
builds a message $\mathsf{info}$ that encodes an identifier 
of the vendor, the number of points this token is worth and an 
expiration date. 
Unlike for purchase receipts, the vendor has no legitimate 
interest in linking
several tokens containing loyalty points; hence, 
the customer picks a fresh random $y$ for each 
new loyalty points token she claims. 
Then the vendor and the
customer run the interactive protocol 
$\mathsf{Issuance}(\vendor, \part, \mathsf{info}, y)$.
The generated token contains the loyalty points issued to the customer.

To ensure that a loyalty points token submitted for redemption
cannot be linked with an issued loyalty points token,
the number of loyalty points associated to 
a single token should be 
limited to a small set of possible values, similar to 
the limited denominations of bank notes.
There is an efficiency toll to be paid for 
this caution, as issuing a certain
amount of loyalty points can require 
running the $\mathsf{Issuance}$ protocol several
times (several tokens may be needed to reach
the required amount).

\subsection{Redeem} 

A participant $\cust$ who wants to redeem a 
loyalty points token $T$ previously earned at a vendor $\vendor$'s 
in exchange for some benefits 
runs the interactive protocol
$\mathsf{Verification}(\vendor, \part, T)$.

It is possible to simultaneously redeem 
several loyalty points tokens 
by using the aggregation
of signatures 
described in Section~\ref{desc}.


\subsection{Incentives related to purchase receipts submission}
\label{reward}

Vendors can establish strategies to incentivize or discourage
certain customer behaviors:
\begin{itemize}
\item To encourage customers to use little or no
purchase receipt generalization (and hence to renounce
some of their privacy), the amount of loyalty points
awarded per receipt token should depend on the chosen
level of generalization: more loyalty points awarded
to less generalized purchase receipts.
\item
If the customer submits unlinkable receipts, she should
just get enough loyalty points to reward her as a returning customer.
To encourage customers to allow linkage of 
purchase receipt tokens by the vendor (and hence
customer profiling),  
a customer should get more loyalty points if she submits 
$n_1 + n_2$ tokens with the same $y$ value than if she submits
$n_1$ tokens with one $y$ value and then $n_2$ tokens with a different
$y$ value ({\em superlinear reward}).
Furthermore, the vendor may require
that  
the list of linkable receipt tokens for which reward is claimed
correspond to purchases made within a certain time
window (if linking purchases very distant in time is uninteresting
for profiling). 
\item Two or more customers might be tempted
to share their $y$ values in order to submit a longer
list of linkable receipts and thereafter share the superlinear
number of loyalty points they would earn. 
As long the reward is only {\em slightly} superlinear,
customer collusion is discouraged if the customer 
$\cust$ who submits the list
of linkable tokens is required by $\vendor$ 
to actually show all the actual linkable tokens
(and not just a reference to them): 
colluders different from $\cust$ may not like 
to pay the privacy toll of disclosing
their purchase receipts to $\cust$.
\end{itemize}



\section{Performance analysis}\label{sec.experiments}
We count here the number of operations required by 
the $\mathsf{Issuance}$ and $\mathsf{Verification}$ protocols described
in Section~\ref{sec.annt}.

The $\mathsf{Issuance}$ protocol requires the
computation by the vendor of 
1 exponentiation in $\G$; also, 
1 hash, 1 addition and 1 inversion 
in $\Zahl_q^*$. The customer computes 
2 exponentiations
in $\G$ and 1 inversion in $\Zahl_q^*$.
The $\mathsf{Verification}$ protocol requires the
computation by the vendor of 
1 exponentiation, 1 multiplication and 1 hash in $\G$; also, 1 hash
in $\Zahl_q^*$ and 2 pairings.

We used the jPBC library~\cite{jPBC} to test
times to compute each of the operations.
We generated a symmetric pairing
constructed on the curve $y^ 2 = x^ 3 + x$ with characteristic
a 512-bit prime and embedding degree 2, {\em i.e.}, the Type A
pairings suggested in~\cite{Lynn2007}. 
The order of $\G$ over the curve is
a prime of 160 bits, elements in $\G$ are 512 bits
long and elements in $\Zahl_q^*$ 160 bits long.

With the above technology choices, a multiplication of points in $\G$ takes 0.09 ms,
an exponentiation in $\G$ takes 17.2 ms,
an exponentiation in $\G$ (with precomputation) takes 2.48 ms,
a pairing takes 20.8 ms, and a 
pairing (with precomputation) takes 10.76 ms.

\section{Conclusions and future work}\label{sec.conclusions}
In our privacy-preserving alternative to traditional loyalty programs,
the customers are granted the power to decide what
private information they want to disclose, and how precise
that information is.
We have described a privacy-preserving 
protocol suite that still offers the two
main features of loyalty programs: reward returning customers
and make customer profiling possible.

Future research will involve hiding the $y$ values 
to technically deter customer collusions in purchase receipt submission.
Also, in the context
of a Google Faculty Research Award that partially
funds this work, we plan to implement our solution using smartphones
on the customer's side and test a demonstrator
to show its practical feasibility.


\section*{Acknowledgments}

We thank Dr Qiang Tang for useful discussions on an earlier
version of this paper.
The following funding sources are acknowledged:
Google (Faculty Research Award to the second author), 
Government of Catalonia (ICREA Acad\`emia Prize to the 
second author and grant 2014 SGR 537),
Spanish Government (project TIN2011-27076-C03-01 ``CO-PRIVACY''),
European Commission (FP7 projects
``DwB'' and ``Inter-Trust'') and Templeton World Charity
Foundation (grant TWCF0095/AB60 ``Co-Utility'').
The authors are with the UNESCO Chair in Data Privacy.
The views in this paper are the authors' own and 
do not necessarily reflect
the views of Google, UNESCO or the Templeton World Charity
Foundation.



\begin{thebibliography}{1}


\bibitem{Abe96}
M.~Abe, E.~Fujisaki, 
``How to date blind signatures,''
\emph{Advances in Cryptology--ASIACRYPT 1996}, pp. 244--251, Springer, 2002.

\bibitem{Boneh2001}
D.~Boneh, M.~Franklin,
``Identity-based encryption from the Weil pairing,''
\emph{Advances in Cryptology--CRYPTO 2001}, pp. 213--229, Springer, 2001.

\bibitem{Camenisch01}
J.~Camenisch, A.~Lysyanskaya,
``An efficient system for non-transferable anonymous credentials 
with optional anonymity revocation,''
\emph{Advances in Cryptology--EUROCRYPT 2001}, pp. 93--118, Springer, 2001.

\bibitem{Camenisch04}
J.~Camenisch, A.~Lysyanskaya,
``Signature schemes and anonymous credentials from bilinear maps,''
\emph{Advances in Cryptology--CRYPTO 2004} pp. 56--72, Springer, 2004.

\bibitem{Chaum83}
D.~Chaum,
``Blind signatures for untraceable payments,''
\emph{Advances in Cryptology--CRYPTO 82}, pp. 199--203, Springer, 1983.

\bibitem{Chaum90}
D.~Chaum, A.~Fiat, M.~Naor,
``Untraceable electronic cash,''
\emph{Advances in Cryptology--CRYPTO 88}, pp. 319--327, Springer, 1990.

\bibitem{Rep3}
``Consumers reveal privacy concerns with loyalty programs''
\emph{Convenience Store Decisions}, 2014. \url{www.csdecisions.com}

\bibitem{jPBC}
A.~De Caro, V.~Iovino,
``jPBC: Java pairing based cryptography,''
\emph{Proceedings of the 16th IEEE Symposium on Computers and Communications},
pp. 850--855, IEEE, 2011.

\bibitem{Rep4}
C.~Dunn,
``Loyalty programs and privacy issues: do you need to worry about providing 
personal information?''
\emph{Disney Family}, accessed \today. \url{family.go.com}


\bibitem{eDropship}
R.C.~Johnson,
``eDropship: methods and systems for anonymous eCommerce shipment,''
\emph{US Patent 7,853,481}, 2010.

\bibitem{Lynn2007}
B.~Lynn,
{\em On the Implementation of Pairing-based Cryptosystems},
Doctoral dissertation, Stanford University, 2007.


\bibitem{Zerocoin}
I.~Miers, C.~Garman, M.~Green, A.~D.~Rubin, 
``Zerocoin: anonymous distributed e-cash from Bitcoin,''
\emph{Security and Privacy (SP), 2013 IEEE Symposium on}, pp. 397--411, IEEE, 2013.

\bibitem{Bitcoin}
S.~Nakamoto,
``Bitcoin: A peer-to-peer electronic cash system,''
\emph{Consulted}, vol. 1, 2008. Available in \url{www.bitcoin.org/bitcoin.pdf}.

\bibitem{Neff01}
C.A.~Neff, (2001, November). 
``A verifiable secret shuffle and its application to e-voting,''
\emph{Proceedings of the 8th ACM conference on Computer and Communications Security}, pp. 116--125, ACM, 2001.

\bibitem{Rep1}
A.~Pratt,
``Loyalty cards vs. privacy concerns,''
\emph{Infosec ISLAND}, 2011. \url{www.infosecisland.com}


\bibitem{ToR}
``Tor Project: Anonymity Online,'' accessed \today. 
\url{www.torproject.org}

\bibitem{Rep2}
B.~Tuttle,
``A disloyalty movement? Supermarkets and customers drop loyalty card programs''
\emph{TIME}, 2013. \url{business.time.com}

\bibitem{PBBLS}
F.~Zhang, R.~Safavi-Naini, W.~Susilo,
``Efficient verifiably encrypted signature and partially blind signature from bilinear pairings,''
\emph{Progress in Cryptology--INDOCRYPT 2003}, pp. 191--204, Springer, 2003.

\end{thebibliography}
\end{document}